\documentstyle[aaspp4,11pt]{article}

\def\simlt{\lower.5ex\hbox{$\; \buildrel < \over \sim \;$}}
\def\simgt{\lower.5ex\hbox{$\; \buildrel > \over \sim \;$}}
\def\gsim{\;\rlap{\lower 2.5pt
\hbox{$\sim$}}\raise 1.5pt\hbox{$>$}\;}
\def\lsim{\;\rlap{\lower 2.5pt
   \hbox{$\sim$}}\raise 1.5pt\hbox{$<$}\;}
\def\msun{{\rm\,M_\odot}}

\def\spose#1{\hbox to 0pt{#1\hss}}
\def\lta{\mathrel{\spose{\lower 3pt\hbox{$\mathchar''218$}}
     \raise 2.0pt\hbox{$\mathchar''13C$}}}
\def\gta{\mathrel{\spose{\lower 3pt\hbox{$\mathchar''218$}}
     \raise 2.0pt\hbox{$\mathchar''13E$}}}
\newcommand{\beq}{\begin{equation}}
\newcommand{\eeq}{\end{equation}}


\lefthead{MENOU \& QUATAERT}
\righthead{{TIDAL DISRUPTION EVENTS}}

\begin{document}

\title{Activity From Tidal Disruptions in Galactic Nuclei}

\author{Kristen Menou\altaffilmark{a,1} \& Eliot Quataert\altaffilmark{b,c,1}}

\affil{$^{a}$Princeton University, Department of Astrophysical Sciences, Princeton, NJ 08544; kristen@astro.princeton.edu}
\affil{$^{b}$Institute for Advanced Study, School of Natural Sciences, Einstein Drive, Princeton, NJ 08540}
\affil{$^{c}$Present Address:  Astronomy Department, UC Berkeley, Berkeley, CA 94720; eliot@astron.berkeley.edu}

\altaffiltext{1}{Chandra Fellow}

\authoremail{kristen@astro.princeton.edu}
\authoremail{eliot@ias.edu}

\vspace{\baselineskip}
%\submitted{Submitted to ApJ Letters}

\begin{abstract}

The tidal disruption of a star by a supermassive black hole is
expected to lead to a short bright flare followed by an extended
period of low-level emission.  Existing models of the late-time
accretion of the stellar debris via a thin disk imply that the
Galactic Center source Sgr A* should be brighter than currently
observed by several orders of magnitude. { A similar problem exists}
for M31 and M32. This problem could be avoided if thin disk accretion
transitions to low radiative efficiency accretion (e.g., ADAF, CDAF)
at low accretion rates (via ``evaporation'' of the thin disk).
Alternatively, we show that the outer portions of a thin disk may
become neutral and unable to sustain MHD turbulence on a timescale
less than the time between tidal disruption events; this may
dramatically shorten the duration of the late accretion phase.

\end{abstract}

{\it subject headings}: accretion, accretion disks -- black hole
physics -- MHD -- turbulence -- galaxies: nuclei 

\section{Introduction}

{ A main-sequence star passing within a few Schwarzschild radii of
a supermassive black hole (SMBH) of mass $M_{\rm bh} \lsim 2 \times
10^8 \msun$ is disrupted by strong tidal forces.  Tidal disruption
events may therefore offer signatures, in normal (non-active)
galaxies, of the presence of a SMBH (e.g. Rees 1990).} In recent
years, our knowledge of the demographics of SMBHs has improved
dramatically, in large part thanks to dynamical studies. There is now
strong evidence for a SMBH at the center of our galaxy (Genzel et
al. 1997; Ghez et al. 2000) as well as in most galaxies possessing a
bulge-component (Magorrian et al. 1998; Gebhardt et al. 2000;
Ferrarese \& Merritt 2000; Gebhardt et al. 2001; Merritt et
al. 2001). This evidence relies on detailed modeling of the stellar
dynamics in galactic nuclei, which also provides reliable estimates of
the stellar disruption rate (Magorrian \& Tremaine 1999; Syer \& Ulmer
1999).

Most work on the observational signatures of tidal disruption events
has focused on flares (typically in the UV) lasting for a few years
and corresponding to the early accretion phase of the stellar debris
bound to the SMBH (Rees 1988; Ulmer 1999). Cannizzo, Lee, \& Goodman
(1990), however, studied the late time accretion of the stellar debris
via a geometrically thin accretion disk. These authors found that the
disk accretion rate decreases slowly as a power-law in time, which
suggests that remnant activity in galactic nuclei may be visible a
long time after the last disruption event occurred. Cannizzo et
al. (1990) noted that their results could be problematic in galaxies
such as M32, since the nucleus should be brighter than is observed
(see also Ulmer 1997).

Here, we revisit the problem of the accretion of stellar debris
following a tidal disruption event by a supermassive black hole. In
\S2, we briefly review some of the basic theory concerning stellar
disruption events. We then point out that the standard accretion
scenario, when combined with the latest disruption rates, implies that
the SMBH in our Galactic Center (Sgr A*) should be brighter than is
observed by several orders of magnitude. In \S3, we reconsider the
accretion scenario, paying particular attention to the structure and
properties of the accretion flow during both the early super-Eddington
phase and the late sub-Eddington phase.  In \S4, we conclude with a
few possible extensions of this work.

\section{Basics of Tidal Disruption}

\subsection{Theory}

A solar-type star is tidally disrupted if its orbit brings it closer
to a SMBH (with $M_{\rm bh} \lsim 2 \times 10^8 \msun$) than the tidal
radius $R_{\rm T} \approx 5 M_7^{-2/3} R_s$, where $M_7 = M_{\rm
bh}/10^7 M_\odot$ and $R_s = 3 \times 10^{12} M_7$~cm is the
Schwarzschild radius of the SMBH (e.g. Hills 1975). Following the
disruption, about half of the stellar debris is bound to the SMBH and
half is unbound.  The bound material is nearly uniformly distributed
in binding energy and so is distributed as $P^{-5/3}$ in orbital
period, with a minimum orbital period $P_{\rm min} \approx 0.7
M_7^{1/2}$~yr (Rees 1988; Evans \& Kochanek 1989).

The subsequent evolution of the bound stellar debris is both difficult
to predict analytically and to follow numerically (Lacy, Townes \&
Hollenbach 1982; Rees 1988; Evans \& Kochanek 1989; Cannizzo et
al. 1990; Kochanek 1994). It is unclear what fraction of the initially
bound material remains bound after the second or subsequent periastron
passages (violent shocks and significant gas heating being expected
from stream-stream collisions). The latest numerical simulations
indicate that about $25\%$ of the initial stellar mass may remain
bound after the second periastron passage (Ayal, Livio \& Piran 2000).

Although the process of circularization by stream-stream collisions is
complex, one expects circularization to occur on a timescale
corresponding to at most a few orbital timescales, $P_{\rm min}$, at
pericenter (Ulmer 1999). Given the distribution of binding energy of
the bound material, mass returns to pericenter at a rate $\dot M
\approx M_7^{-1/2} (t/P_{\rm min})^{-5/3} M_\odot$ yr$^{-1}$ for $t
\gsim P_{\rm min}$ (Rees 1988; Phinney 1989).  This is larger than the
Eddington accretion rate $\approx 0.1 M_7 M_\odot$ yr$^{-1}$ for
several years and so accretion should initially proceed via a photon
trapped radiatively inefficient accretion flow (Rees 1988; Ulmer 1997;
see \S3).  The viscous timescale in such a hot flow is indeed short
enough for the material to accrete as fast as it circularizes (unless
the viscosity parameter $\alpha$ is $\ll 0.1$; Ulmer 1999).

After a few years, the infall rate drops below the Eddington rate,
photons are no longer trapped in the accretion flow, and the gas
begins to cool efficiently.  Mass returning to pericenter then begins
to accrete via a geometrically thin disk, with the accretion rate set
by the long viscous time in the thin disk, not the rate at which mass
returns to pericenter.  Cannizzo et al. (1990) studied the thin disk
accretion phase using numerical simulations and analytic self-similar
solutions.  { They assumed that} the accretion disk has nearly
constant total angular momentum; viscosity can redistribute angular
momentum (from material at small radii to material at large radii),
but it is assumed that there is no global loss of angular momentum
from the system such as would be provided by a jet or an MHD wind.
Under these assumptions, Cannizzo et al. found that the accretion rate
onto the SMBH decays as a weak power-law in time: $\dot M \propto
t^{-n}$, with $n \simeq 1.2$ for various opacity laws and almost
independent of the value of the viscosity parameter $\alpha$ in the
disk.  As the disk mass decreases due to accretion, the mean radius of
the disk increases in order to conserve angular momentum (roughly as
$R(t) \propto t^{3/8}$).

\subsection{Observational Constraints}

Attempts to identify tidal disruption events have mostly focused on
detecting the initial ``flare'' of emission with a timescale of
$\approx P_{\rm min} \approx$ years; { these are short-lived events
with a very low duty-cycle per galaxy of $\sim 10^{-4}$ (e.g. Rees
1988; Loeb \& Ulmer 1997; Ulmer, Paczynski \& Goodman 1998; Komossa \&
Dahlem 2001).}  Another approach to the problem is to look for the
late emission from tidal disruption events, as expected from the study
of disk accretion by Cannizzo et al. (1990). In this scenario, the
issue is not one of short duration, but rather of the detectability of
low-level remnant activity in a galactic nucleus. To address this
question quantitatively, we searched the literature for a set of
nearby galaxies meeting the following three criteria: (1) estimates of
tidal disruption rates based on recent dynamical studies are available
(Magorrian \& Tremaine 1999; Syer \& Ulmer 1999); (2) the estimated
SMBH mass is below $2 \times 10^8 M_\odot$, the maximum mass allowed
for main-sequence stellar disruption; (3) detections of, or useful
upper limits on, low-level nuclear activity are available. The three
galaxies that met these criteria are our own Galaxy (i.e. Sgr A*), M31
and M32.

Table~\ref{tab:one} summarizes the relevant data for these three
galaxies. For the Galactic Center main-sequence disruption rate, we
adopt $10^{-4}$~yr$^{-1}$, which is somewhat in excess of the value
quoted by Syer \& Ulmer (1999). This is because these authors tend to
underestimate the rates as compared to Magorrian \& Tremaine
(1999).\footnote{Presumably, this is due to the additional stellar
orbits considered by Magorrian \& Tremaine in their calculation.}  For
the bolometric luminosities, we conservatively adopt the highest
luminosity value available for each nucleus among the various observed
spectral bands (when detected).

{ For thin disk accretion, the luminosity limits in
Table~\ref{tab:one} can be converted into accretion rate limits for a
radiative efficiency $\approx 10 \%$.}  We find that the accretion
rates onto the SMBHs in Sgr A*, M31, \& M32 would be $\approx
10^{16}$, $< 10^{17}$ and $ < 5 \times 10^{16}$~g~s$^{-1}$,
respectively, for radiatively efficient accretion. These accretion
rates are {\it very small} in the context of tidal disruption theory:
for $\alpha = 0.1$ and an initial $\approx 0.1 M_\odot$ disk of
tidally disrupted material (see \S3), Cannizzo et al.'s (1990) thin
disk models predict that the accretion rate onto the SMBH drops below
$10^{19}$~g~s$^{-1}$ approximately $4 \times 10^4$~yrs { following a
tidal disruption event (see their Fig.~3b), while it reaches values}
$\leq 10^{17}$~g~s$^{-1}$ only well after $10^6$ years.  We expect
these predictions to be relatively robust given the very weak
dependence of $\dot M(t)$ on viscosity and opacity.  Since $10^6$
years is much longer than the expected time between two consecutive
stellar disruptions, disk accretion of stellar debris as described by
Cannizzo et al. (1990) predicts that Sgr A*, M31, and M32 should be
{\it continuously brighter} than is presently observed. {This is
illustrated in Table~\ref{tab:one}, where we also list, for the three
galactic nucei of interest, the luminosity expected in Cannizzo et
al.'s thin disk models after a time corresponding to the interval
between successive disruption events. Note that} our analysis is
conservative since we do not consider the emission spectrum expected
from thin accretion disks { (expected to be prominent in the IR-UV,
where the observational limits are generally stronger than our adopted
bolometric luminosities).}

\section{Revisions to the Accretion Model for Tidal Disruption Events}

One possible resolution of this observational puzzle is the following:
for accretion rates below $\approx 1 \%$ of the Eddington accretion
rate, i.e., $\lsim 10^{-3} M_7 M_\odot$ yr$^{-1}$, gas can accrete
onto a black hole via a hot optically thin low radiative efficiency
accretion flow (LRAF) such as an advection-dominated accretion flow
(ADAF; Rees et al. 1982; Abramowicz et al. 1995; Narayan \& Yi 1994;
1995) or its variants, ADIOSs (Blandford \& Begelman 1999) and CDAFs
(Quataert \& Gruzinov 2000; Narayan et al. 2000).  Typically, an
accretion rate $\dot M \approx 10^{-2} \dot M_{\rm Edd}$ is reached
years to decades after a tidal disruption event.  If thin disk
accretion transitions to LRAF accretion at this time, the accretion
luminosity will precipitously drop, either because the radiative
efficiency decreases dramatically (ADAFs) or because the accretion
rate decreases dramatically (CDAFs/ADIOS), or perhaps both.  This
scenario may provide { an explanation} for the absence of late time
emission from the accretion of tidally disrupted stars (see also Ulmer
1997).

However, the thin disk $\rightarrow$ LRAF model requires positing that
thin disks spontaneously ``evaporate'' into LRAFs when $\dot M \lsim
10^{-2} \dot M_{\rm Edd}$ (thin disks remain, after all, a viable mode
of accretion even for $\dot M \ll \dot M_{\rm edd}$).  This
possibility has been extensively and successfully applied to
observations of time-dependent accretion flows in X-ray binaries {
(e.g., Esin et al. 1997; 1998; 2001),} but the ``evaporation'' process
-- and hence its dependence on, e.g., black hole mass -- { is not
well understood.  It therefore seems} worthwhile exploring scenarios
in which accretion continues to proceed via a thin disk even for $\dot
M \ll \dot M_{\rm Edd}$.

\subsection{Early Super-Eddington Phase}

The initial accretion phase of tidal disruption events are likely to
proceed via a super-Eddington accretion flow in which the radiative
efficiency is low because photons are trapped in the accretion flow
(see \S2.1).  Recent work on such LRAFs indicates that much of the
mass in the accretion disk may be lost to an outflow.  Blandford \&
Begelman (1999) proposed this based on analytic arguments (see also
Narayan \& Yi 1994).  In addition, numerical simulations of LRAFs have
found that the accretion flow structure adjusts so as to transport a
significant flux of energy ($\sim 0.01 \dot M c^2$) from small to
large radii (e.g., Stone, Pringle, \& Begelman 1999; Igumenshchev \&
Abramowicz 1999, 2000); this is the CDAF regime described analytically
by Quataert \& Gruzinov (2000) and Narayan et al. (2000).  Although
the fate of this outward energy flux is not specified in either the
analytic calculations or the numerical simulations, it seems very
likely that this energy drives a strong outflow from the outer edge of
the accretion flow { (see also Hawley, Balbus \& Stone 2001).}

The initial super-Eddington phase is important in the context of
tidally disrupted stars because it sets the initial conditions for the
subsequent thin disk evolution.  In particular, rather than having the
entire bound stellar debris available to accrete, we expect most of
this to be blown away in a strong outflow.  Only the material that
returns to pericenter at significantly sub-Eddington rates remains
bound to the black hole.  Assuming that $0.25 \msun$ of debris remains
bound to the hole after second periastron passage and that essentially
all the mass that would be accreted during the super-Eddington phase
is instead lost to infinity, we deduce from \S2.1 that the mass of
bound material left to accrete at sub-Eddington rates is $\lsim 0.1
M_7^{3/5} M_\odot$.

\subsection{Late Sub-Eddington Phase}

For the $\sim 0.1 M_\odot$ accreted at a sub-Eddington rate, the flow
will adopt a thin disk configuration and the solution described by
Cannizzo et al. (1990) should initially apply. As the accretion disk
spreads and cools down, however, the gas eventually reaches
temperatures $\lsim 10^4$ K for which the disk is subject to the
thermal ionization instability caused by hydrogen recombination (Meyer
\& Meyer-Hofmeister 1981; Lin \& Shields 1986; Cannizzo 1992;
Siemiginovska, Czerny \& Kostyunin 1996).  { Following Menou \&
Quataert (2001), we estimate that the effect of the ionization
instability should be limited to the outermost regions of tidal debris
disks and will not lead to substantial changes in the disk's global
evolution.  As the disk spreads somewhat further, however, it may
eventually become so cool ($\sim 1000-2000$ K) and neutral that MHD
turbulence cannot be sustained.}

MHD turbulence and gravitational instability are presently the only
robust candidates for angular momentum transport in isolated accretion
disks (e.g., Balbus \& Hawley 1998; Gammie 2001). Since tidal debris
disks have very low masses they are probably gravitationally stable.
{ If the outermost regions of a tidal debris disk become passive
when MHD turbulence ceases to operate,} the disk's viscous evolution
will be strongly modified. { Active accretion will proceed only in
regions of the disk with $T_c \gsim 1000$~K, which are confined to
smaller and smaller radii as $\dot M$ decreases with time.}

To quantitatively estimate when the disk's viscous evolution is
modified, we note that a disk annulus becomes thermally unstable when
the accretion rate locally drops below the value $\dot M_{\rm crit}
\left( R \right) \simeq 6 \times 10^{17} M_7^{-0.9} R_{13}^{2.7}~{\rm
g~s}^{-1}$, where $R_{13}$ is the radius of interest in units of
$10^{13}$~cm (see, e.g., Hameury et al. 1998).  After an initial
transient phase of duration $t_0$ (tens of years, set by the viscous
time in the thin disk), the disk accretion rate and the mean radius of
the disk evolve roughly as $\dot M \propto t^{-1.2}$ and $R(t)
\propto t^{3/8}$, respectively (\S2.1), so that $\dot M/R^{2.7} \propto
t^{-2.2}$. Combining this scaling with the one for the critical rate,
$\dot M_{\rm crit}$, we find that the disk's outer regions become
subject to the thermal ionization instability at a time
\begin{equation}
t_n=t_0 \left[ \frac{\dot M(t_0)}{6 \times 10^{17}~{\rm g~s^{-1}}}
\frac{M_7^{0.9}}{R_{13}^{2.7} (t_0)}\right]^{1 \over 2.2},
\label{tn}
\end{equation}
where $\dot M(t_0)$ and $R_{13} (t_0)$ are the values of the disk
accretion rate and disk outer radius at time $t_0$, respectively.

For an initial disk mass of $M_d \approx 0.1 \msun$, $\dot M(t_0) \sim
M_d/t_0$, since a reasonable fraction of the disk's mass is accreted
in the initial few viscous times.  Moreover, the initial outer radius
of the disk is of order the tidal disruption radius, i.e., $R(t_0)
\approx R_T \approx 2 \times 10^{13} M_7^{1/3}$~cm.  Thus,
\begin{equation} t_n
\approx 3 \times 10^3 \left(t_0 \over 10 \ {\rm yrs}\right)^{0.55}
\left(M_d \over 0.1 M_\odot\right)^{0.45} \left(R(t_0) \over
R_T\right)^{-1.2} \ {\rm years}.
\label{tnfinal} 
\end{equation}  
For $R(t_0)$ closer to $10^{14}$~cm, as suggested by Cannizzo et al.'s
simulations (because the disk spreads away from the tidal disruption
radius during the transient accretion phase, $t_0$), the above
estimate for $t_n$ is reduced by a factor of $\approx 5$, to $\lsim
10^3$ years.  This { is smaller} than the typical time between
tidal disruption events in our Galaxy, M31 and M32, implying that a
single power law will not adequately describe the late time evolution
of tidal debris disks in galactic nuclei.

{ Lipunova \& Shakura (2000) have shown} that thin $\alpha$--disks
with fixed outer radii evolve according to $\dot M \propto t^{-3.3}$
and $\dot M \propto t^{-2.5}$ for free-free and electron scattering
opacity, respectively.  The evolution is much faster than a disk which
conserves angular momentum ($\dot M \propto t^{-1.2}$) because the
disk is not forced to spread to larger radii where the viscous time is
longer.  In our problem, the radius bounding the active region shrinks
with time and so the accretion rate will decrease even more rapidly
for $t > t_n$.

These results imply that once the outer regions of the disk become
cold enough, the accretion rate and accretion luminosity from remnant
tidal debris disks will precipitously drop.  It is unclear, however,
whether this is quantitatively sufficient to account for the
observational constraints in \S2.1.  In particular, assuming the sharp
drop in $\dot M(t)$ happens a time $\sim t_n$ after a stellar
disruption, the disk accretion rate still has to decrease by { 3-4}
orders of magnitude from $\dot M(t_n) \sim 10^{20}$~g~s$^{-1}$ to
satisfy the limits on bolometric luminosities given in
Table~\ref{tab:one} (disk accretion being radiatively
efficient). Given that the ratio of the time between two consecutive
disruption events and $t_n$ { is as small as $\sim 10$} for the
three galaxies of interest, this implies that a power law evolution of
the disk accretion rate steeper than $t^{-4}$ is required to just
satisfy the observational limits. If one further requires these nuclei
to be as dim as currently observed for most of the time between
consecutive disruption events, the power law has to be even steeper
than this.

\section{Conclusion}

We have reconsidered the accretion of stellar debris following tidal
disruption by a supermassive black hole. We argued that a large
fraction of the mass initially bound to the black hole will be lost to
infinity via an outflow during the early super-Eddington accretion
phase.  This limits the mass available for late time accretion via a
thin disk to $\sim 0.1 M_\odot$.  We also suggest that, on a timescale
of $\sim 10^3$ years, the thin disk accretion rate will decrease much
more quickly than in the standard similarity solution of Cannizzo et
al. (1990) -- this is because the disk has spread so far from the
black hole that its outer regions are cold, mostly neutral, and may be
unable to sustain MHD turbulence.  Quantitatively, however, it is
unclear if these two effects can explain the very low level of
activity observed in the nuclei of our Galaxy, M31 and M32.

{There are a number of uncertain elements in this evolutionary
scenario that require clarification before a firm statement can be
made on the level of activity expected after tidal disruptions in
galactic nuclei. Perhaps paramount among them is whether thin disks
evaporate into low-radiative efficiency accretion flows (LRAFs) once
the accretion rate is sufficiently sub-Eddington.  For purely thin
disk accretion scenarios, a detailed numerical investigation of the
evolution of a disk with an actively accreting zone that shrinks with
time, as described in \S3.2, would be particularly useful. Note that
there are also additional effects that could further reduce the level
of activity expected, such as significant angular momentum losses to a
jet or an MHD wind during the evolution.}

\section*{Acknowledgments}

Support for this work was provided by NASA through Chandra Fellowship
grant PF9-10006 (to KM) and PF9-10008 (to EQ) awarded by the
Smithsonian Astrophysical Observatory for NASA under contract
NAS8-39073. KM thanks the Center for Astrophysical Sciences at Johns
Hopkins University for hospitality.

\clearpage

\begin{table*}
\caption{NEARBY GALACTIC NUCLEI}
\begin{center}
\begin{tabular}{ccccc} \hline \hline
\\
 & Black Hole Mass & Stellar Disruption Rate & Bolometric Luminosity & Expected Luminosity$^a$\\
& ($\msun$) & (yr$^{-1}$)& (erg~s$^{-1}$) & (erg~s$^{-1}$)\\
\\
\hline
\\
Milky Way & $2.6 \times 10^6$~$^b$& $10^{-4}$~$^c$& $10^{36}$~$^e$ & $\sim 5 \times 10^{39}$\\
M31 & $3.5 \times 10^7$~$^b$&  $10^{-4.68}$~$^d$& $< 10^{37}$~$^f$ & $\sim 10^{39}$\\
M32 & $3.7 \times 10^6$~$^b$&  $10^{-3.62}$~$^d$& $< 5 \times 10^{36}$~$^g$ & $\sim 10^{40}$\\
\\
\hline
\end{tabular}
\label{tab:one}
\end{center}
NOTE. -- (a) Adapted from Cannizzo et al. (1990) (b) Gebhardt et
al. (2000) (c) Adapted from Syer \& Ulmer (1999) (d) Magorrian \&
Tremaine (1999) (e) Broadband: Narayan et al. (1998); X-ray: Baganoff
et al. (2001) (f) Radio: Gregory \& Condon (1991); X-ray: Garcia et
al. (2001) (g) Far-IR: Knapp et al. (1989); Steady X-ray: Eskridge,
White \& Davis (1996), Loewenstein et al. (1998).
               
\end{table*}

\end{document}